\documentclass[12pt,preprint]{aastex}
\newcommand{\masyr}{mas\,yr$^{-1}$}
\newcommand{\mura}{$\mu_{\alpha *}$}
\newcommand{\mudec}{$\mu_{\delta}$}
\newcommand{\kms}{km\,s$^{-1}$}

\newcommand{\ebv}{$E_{B-V}$}
\newcommand{\evj}{$E_{V-J}$}
\newcommand{\evh}{$E_{V-H}$}
\newcommand{\evk}{$E_{V-K_s}$}
\newcommand{\rv}{$R_V$}
\newcommand{\av}{$A_V$}

\newcommand{\msun}{$M_{\odot}$}
\newcommand{\bvo}{($B-V$)$_0$}
\newcommand{\mv}{$M_V$}

\slugcomment{To appear in November 2006 {\it Astronomical Journal}}
\shorttitle{New Candidate Cluster}
\shortauthors{Mamajek}
\begin{document}

\title{A New Nearby Candidate Star Cluster in Ophiuchus at d $\simeq$
170\,pc}

\author{Eric E. Mamajek\altaffilmark{1}}
\affil{Harvard-Smithsonian Center for Astrophysics, 60 Garden
St., MS-42, Cambridge, MA, 02138}
\altaffiltext{1}{Clay Postdoctoral Fellow}

\begin{abstract} The recent discoveries of nearby star clusters and
associations within a few hundred pc of the Sun, as well as the order
of magnitude difference in the formation rates of the embedded and
open cluster populations, suggests that additional poor stellar groups
are likely to be found at surprisingly close distances to the Sun.
Here I describe a new nearby stellar aggregate found by virtue of the
parallel proper motions, similar trigonometric parallaxes, and
consistent color-magnitude distribution of its early-type members.
The 120 Myr-old group lies in Ophiuchus at $d$ $\simeq$ 170\,pc, with
its most massive member being the 4th-magnitude post-MS B8II-III star
$\mu$ Oph.  The group may have escaped previous notice due to its
non-negligible extinction ($A_V$ $\simeq$ 0.9\,mag). If the group was
born with a normal initial mass function, and the nine B- and A-type
systems represent a complete system of intermediate-mass stars, then
the original population was probably of order $\sim$200 systems. The
age and space motion of the new cluster are very similar to those of
the Pleiades, $\alpha$ Per cluster, and AB Dor Moving Group,
suggesting that these aggregates may have formed in the same
star-forming complex some $\sim$10$^8$ yr ago.
\end{abstract}


\section{Motivation \label{motivation}}

This is the first in a series of papers regarding the identification
and characterization of groups of young stars within a few hundred pc
of the Sun (mostly, newly discovered groups of young stars).  The
second paper (Mamajek, in prep.) will detail the identification of a
new $\sim$20-Myr-old group within 100\,pc. In this contribution, I
discuss a new candidate group situated 170\,pc away in Ophiuchus that
appears to be similar in age to the Pleiades ($\approx$120 Myr).

Very young stars with ages of $\lesssim$3\,Myr are usually found in
embedded clusters (ECs) with tens to hundreds of members, and
associated with dark molecular clouds
\citep[e.g.][]{Lada03,Porras03}. The vast majority of embedded
clusters must not remain as bound structures identifiable as open
clusters (OCs) after their molecular gas is removed.  This can be
inferred by estimating the formation rates of the local samples of ECs
and OCs within 1 kpc of the Sun.  In comparing the number of ECs to
OCs within the same representative volume, and assuming a constant
cluster-formation rate, \citet{Lada03} calculate that
$\lesssim$4-7\%\, of embedded star clusters probably survive to the
age of the Pleiades ($\approx$130 Myr). Counting only those ECs with
$\gtrsim$35 members from the catalogs of \citet{Lada03} and
\citet{Porras03} within 1 kpc of the Sun, the local surface density of
ECs in the Galactic disk appears to be $\sim$9 kpc$^{-2}$. The number
of ECs with 10-35 members is roughly equal to the number of ECs with
$>$35 members \citep{Porras03}, so the surface density is doubled if
one counts groups with $>$10 members.  Assuming a flat
cluster-formation rate, and a mean EC age of $\sim$2\,Myr,
\citet{Lada03} estimate for their sample of local ECs with $>$35
members a formation rate of $\sim$4 Myr$^{-1}$ kpc$^{-2}$.  If one
assumes that the census of OCs with ages of 10-100\,Myr and $d$ $\leq$
1\,kpc is complete \citep[using the 2006
update\footnote{http://www.astro.iag.usp.br/$\sim$wilton/} of the
catalog of ][]{Dias02}, then the local density of such OCs is
$\sim$35\,kpc$^{-2}$, and the OC formation rate is $\sim$0.34
Myr$^{-1}$ kpc$^{-2}$ (similar to previous estimates by
\citet{Elmegreen85} and \citet{Battinelli91}). The actual cluster
formation rate may be somewhat higher as nearby clusters are still
being discovered \citep{Platais98,Mamajek99,Alessi03,Kharchenko05}.
These calculations suggest that only $\sim$9\% of the ECs seen today
would be hypothetically identified as OCs $\sim$10-100 Myr in the
future, and that $\sim$90\% of ECs likely evolve into unbound
structures that might be identifiable as ``associations'' or ``moving
groups''.  Another conclusion is that for every cataloged OC in the
Sun's vicinity with age $<$100\,Myr, there may be an {\it order of
magnitude} more stellar aggregates with $>$35 members that are {\it
not} cataloged as OCs, and there are likely to be approximately {\it
double} that number again if one tracks the evolution of ECs with
$>$10 members. So where are these fossil remnants of ECs that are
predicted to be in copious supply?

The nearest unbound remnants of ECs that contain any O-type stars,
and/or significant numbers of B-type stars, have probably been
identified as OB associations (OBAs), at least up until ages of
few$\times$10 Myr \citep{Briceno06}.  The progenitors of OBAs were
probably giant molecular cloud complexes containing ensembles of ECs
(with the evolutionary descendants of the individual ECs appearing
later as ``subgroups'' in the OBAs). The OB associations within 1 kpc
cataloged by \citet{deZeeuw99} contain between 6 and 85 B-type stars
($\overline{n_B} \, \simeq\, 45$).  Young stellar groups with ages of
$\sim$3-100 Myr that contain few, if any, B-type stars have only recently
been identified. Given a standard initial mass function (IMF), one
naively expects that for each B-type star formed, an EC will also
typically produce $\sim$1 A-star, $\sim$1 F-star, $\sim$1.5 G-stars,
$\sim$4 K-stars, and $\sim$28 M-stars \citep[where I have adopted the
IMF of ][ and the spectral types are reflective of the resultant main
sequence population]{Kroupa01}. Hence, if born with a \citet{Kroupa01}
IMF, an unrecognized cluster or association with $\sim$5 B-type stars
may have represented an embedded cluster of $\sim$200\,stars.  That
such small groups can escape detection is well illustrated by that
fact that at least 5 well-studied stellar aggregates with ages of
$\lesssim$30\,Myr and distances of $\lesssim$110 pc have been
discovered over the past decade \citep[e.g. $\eta$ Cha, $\epsilon$
Cha, TW Hya, $\beta$ Pic, Tuc-Hor
groups;][]{Mamajek99,Mamajek00,Feigelson03,Kastner97,
Barrado99,Zuckerman00,Torres00}.  Each contains between zero (TW Hya
Assn.) and four (Tuc-Hor) B-type stars, tens of low-mass stars, and no
molecular gas (except for $\epsilon$ Cha, the youngest of
these). Lacking bright, hot stars from their ranks, the discovery of
these groups had to await the era of large uniform astrometric
databases with accurate proper motions and trigonometric parallaxes
\citep[i.e. {\it Hipparcos};][]{Perryman97} and an X-ray all-sky
survey \citep[i.e. {\it ROSAT};][]{Voges99}.  Any heretofore missing
groups may be identifiable by virtue of their common proper motions
and distances (where the higher mass stars will be preferentially
conspicuous due to the shallow magnitude limits of contemporary
astrometric catalogs).  The low-mass members of such groups may be
identifiable by virtue of various youth signatures, with strong X-ray
emission being the most conspicuous. The most unbound aggregates may
even lack a distinguishable nucleus (e.g. TW Hya association), and may
simply be identifiable as a slight over-density of young stars of
similar age and space motion.

\section{Analysis \label{analysis}}

In an effort to further investigate the memberships and kinematics of
known young, nearby stellar associations, as well as identify new
ones, the author has compiled an extensive list of astrometric data
for a sample of plausibly young stars within a few hundred pc of the
Sun. This list includes OBA-type stars from the compiled ASCC
astrometric catalog \citep{Kharchenko01}, {\it ROSAT} All-Sky Survey
X-ray sources \citep[RASS;][]{Voges99,Voges00} with optical
counterparts in the Tycho-2 \citep{Hog00} or UCAC2 \citep{Zacharias04}
catalogs, and previously identified pre-MS stars
\citep{Ducourant05,Herbig88}. For constructing the RASS/Tycho-2/UCAC2
catalog, an optimal X-ray/optical separation of 40'' was adopted
\citep{Neuhauser95}.  This compilation of catalogs provides critical
astrometric and X-ray data for characterizing the membership and
kinematics of nearby, young stellar groups, as well as identifying new
ones.

\begin{figure}
\plotone{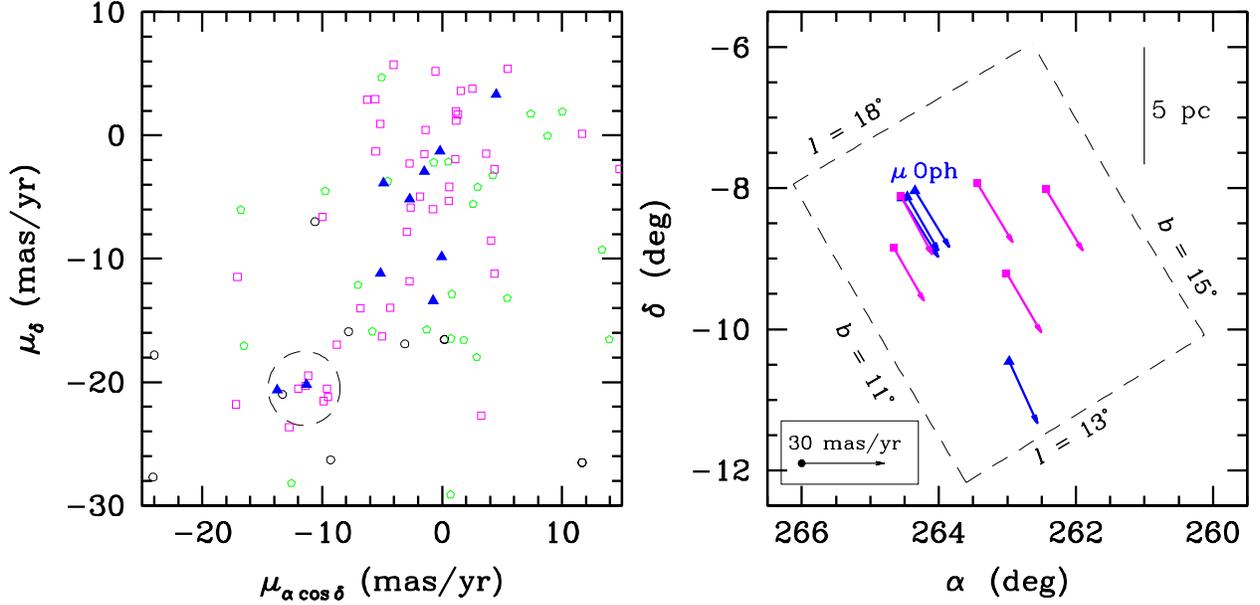}
\caption{Proper motions of stars in the vicinity of $\mu$ Oph. Colored
points are stars from the ASCC-2.5 catalog, representing stars of
spectral type B ({\it filled blue triangles}), A ({\it open magenta
squares}), and F ({\it open green pentagons}). Only ASCC stars with
measured parallaxes, $BV$ magnitudes, and spectral types are
plotted. X-ray stars from the {\it ROSAT} All Sky Survey with
counterparts in the UCAC2 and Tycho-2 catalogs are shown as {\it open
black circles}. Some ASCC stars appear twice (i.e. they are RASS X-ray
sources). HD 160038 (B9V) is the only X-ray source selected.  The {\it
left plot} shows the proper motions for ASCC and RASS stars over the
entire 7$^{\circ}$ $\times$ 7$^{\circ}$ region centered on ($\alpha$,
$\delta$) = (263$^{\circ}$.8, --9$^{\circ}$.0).  The {\it right plot}
shows the positions and proper motion vectors for stars with proper
motions encircled in the left plot (within 3 \masyr\, of \mura,
\mudec\, = --11.5, --20.5 \masyr). The {\it dashed lines} show
Galactic latitude and longitude lines which encompass the group.  The
5\,pc segment is for the adopted distance (173\,pc; \S\ref{distance}).
\label{fig:muoph_vector}}
\end{figure}

\begin{figure}
\epsscale{0.6}
\plotone{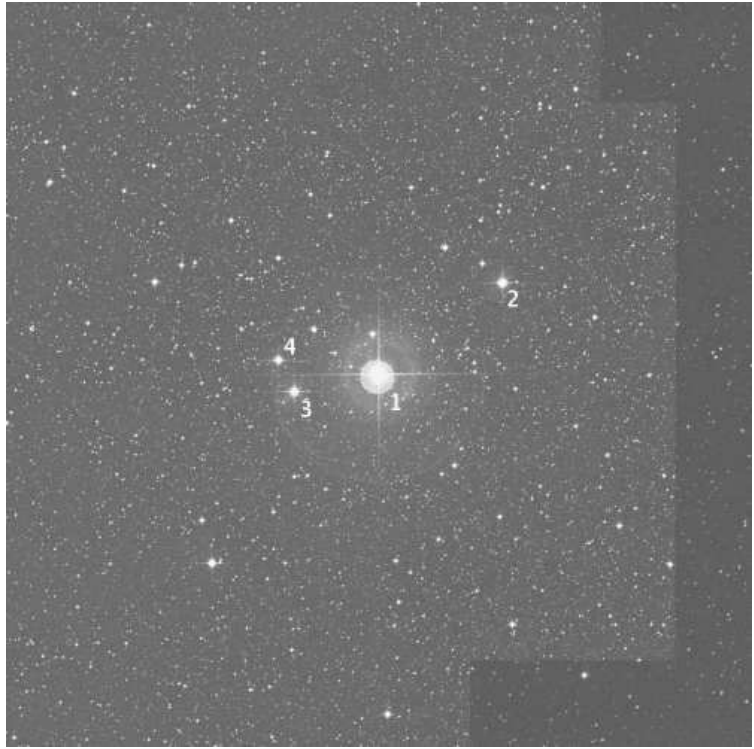}
\caption{
Second generation Digitized Sky Survey (red) image centered on $\mu$
Oph.  Field of view is 40' (2.0\,pc at $d$ = 173\,pc). Stars are
indexed in Table \ref{tab:muoph_members}, where \#1 is $\mu$ Oph itself.
\label{fig:muoph_dss}}
\end{figure}

The Ophiuchus region is rich in recent, and on-going, star-formation
at $d$ $\simeq$ 140\,pc \citep[e.g. LDN 1688, Upper Sco subgroup of
Sco-Cen OB association, etc.;][]{Blaauw91,deZeeuw99,Preibisch06}.  In
plotting the positions and proper motion vectors for young stars over
several thousand square degrees in the general direction of the
Oph-Sco-Cen star-formation region, I noticed a previously
unrecognized, tight group of early-type stars with nearly parallel
proper motions in Ophiuchus (\mura, \mudec\, $\simeq$\, --11.5, --20.5
\masyr; Fig.~\ref{fig:muoph_vector}).  The densest concentration of
stars in this group is roughly centered on $\mu$ Oph
\citep[B8II-III;][]{Houk99}, which appears to be the brightest member
of the group \citep[$V$ = 4.58$^m$;][]{Perryman97}. Several early-type
stars with ASCC proper motions nearly identical to that of $\mu$ Oph
can be found to the south and west of the $\mu$ Oph clump, but no
obvious additional early-type candidate members to the north and east
were found. Four early-type stars within a $\sim$20' diameter region
near ($\alpha, \delta$ = 264$^{\circ}$.5, --8$^{\circ}$.1;
17$^h$38$^m$ --8$^{\circ}$6'; J2000) comprise what appears to be the
``nucleus'' of this group: $\mu$ Oph (HD 159975), HD 159874 (B8IV/V;
$V$ = 7.83$^m$), HD 160038 (B9V; $V$ = 7.98$^m$), and HD 160037
\citep[A0V; $V$ = 8.90$^m$;][]{Houk99,Perryman97}. The projected
separations between these four stars (at an assumed distance of
173\,pc; \S\ref{distance}) are 0.1-0.7\,pc, which are larger than the
maximum observed separations for binary stars with late B-type
primaries \citep[$\sim$0.1\,pc; e.g.][]{Abt88}. A Digitized Sky Survey
image of the nucleus region is shown in Fig. \ref{fig:muoph_dss}, and
the nucleus stars are indexed 1-4 in order of brightness at $V$ band.

\subsection{Selection \label{selection}}

I assembled a preliminary sample of candidate members by selecting
those early-type stars with ASCC proper motions within 3\,\masyr\, of
(\mura, \mudec) = (--11.5, --20.5 \masyr), and with positions within a
7$^{\circ}$ box centered on ($\alpha, \delta$ = 263$^{\circ}$.8,
--9$^{\circ}$.0; J2000). The motivation for using these particular
selection values is empirical, based on the conspicuous presence of
the ``nucleus'' in position and proper motion space
(Figs. \ref{fig:muoph_vector} \& \ref{fig:muoph_dss}), and the clean
detachment of the group from other stars in proper motion space.  Note
that for an assumed distance of 173\,pc (see \S\ref{distance}),
1\,\masyr\, translates into 0.8\,\kms.

The astrometric data for these stars are provided in Table
\ref{tab:muoph_astrom}, and the photometric data and spectral types
are presented in Table \ref{tab:muoph_members}.  The points on the
color-magnitude diagram for these objects are roughly consistent with
constituting a reddened, co-distant population (discussed further in
\S\ref{age}).  The hypothesis that this is a stellar association of
some kind seems to be tenable, so we proceed to solve for the
reddening, age, and distance of the system.

\subsection{Reddening \label{reddening}}

There are two good reasons for investigating the reddening of the
$\mu$ Oph group candidate members in more detail: (1) one needs to
account for reddening in order to estimate the intrinsic brightnesses
of the stars; critical to estimating the isochronal age of the
aggregate, and (2) there are reports in the literature of anomalous
reddening in the general direction of Oph and Sco
\citep{Whittet80,Turner89}, in which case the standard ratio of total
to selective extinction ($R_V$ = $A_V$/$E_{B-V}$) might not be close
to the (commonly assumed) Galactic mean value of $R$ $\simeq$ 3.1
\citep{Mathis90}. Using the photometry and spectral type data in
Table \ref{tab:muoph_members}, and eqn. 39 and the 2MASS and Johnson
filter data from \citet[][for the regime of hot stars with low
reddening]{Fiorucci03}, I estimated $R_V$ via three derived equations:

\begin{equation}
R_V\,=\,1.69\frac{E(V-J)}{E(B-V)} - 0.66
\end{equation}

\begin{equation}
R_V\,=\,1.35\frac{E(V-H)}{E(B-V)} - 0.34
\end{equation}

\begin{equation}
R_V\,=\,1.20\frac{E(V-K_s)}{E(B-V)} - 0.18
\end{equation}

For a given star, the three \rv\, values are not independent
estimates, as they all depend on the same $B$ and $V$ photometry. For
this reason, I quote in Table
\ref{tab:muoph_members} the mean \rv\, values and adopt their standard
deviation (rather than the standard error) as an estimate of the
uncertainty in \rv. An additional (negligible) uncertainty of $\pm$0.1 in
\rv\, was included to take into account the uncertainty in the spectral
types (assumed to be accurate to $\pm$1 subtype).  The typical
uncertainty in \rv\, for a given star was then $\pm$0.2.  The
unweighted mean \rv\, value among the candidate members is
$\overline{R_V}$ = 3.53\,$\pm$\,0.16 ($\sigma$ = 0.45), after
excluding star \#9 (HD 158875; \rv\,=\,6.1), which fails Chauvenet's
criterion and can be considered a statistical outlier
\citep{Bevington92}.  As the 2MASS photometry for $\mu$ Oph itself was
of poor quality, no useful estimate of \rv\, could be calculated, so I
adopt the mean group value for this star. That the dispersion in
\rv\, values is larger than the observational uncertainties suggests
that it is more appropriate to adopt the individual \rv\, values in
Table \ref{tab:muoph_members} for calculating extinctions, rather than
adopting a mean group value. The mean \rv\, (3.5) is intermediate
between the Galactic mean value (3.1) and that reported for the
Sco-Oph region by \citet[][\rv\, =\, 3.9]{Whittet80}.  The \bv\, color
excesses for the nine stars in Table \ref{tab:muoph_members} range
from \ebv\, = 0.19 to 0.31, and are consistent with a mean value of
$\overline{E_{B-V}}$ = 0.26\, $\pm$\, 0.02 ($\sigma$ = 0.05 mag). The
mean $V$ band extinction for the B- and A-type stars is then
$\overline{A_V}$ $\simeq$ $\overline{R_V}$ $\times$
$\overline{E_{B-V}}$ = 0.9\,$\pm$\,0.1 mag.

\subsection{Distance \label{distance}}

The {\it Hipparcos} and Tycho parallaxes for the 9 group members in
Table \ref{tab:muoph_astrom} are statistically consistent with one
another, within their published uncertainties. The
inverse-variance-weighted mean trigonometric parallax for the sample
is $\overline{\varpi}$\, =\, 5.77\, $\pm$\, 0.48\, mas, which
translates to a distance of 173$^{+16}_{-13}$ pc.  For the 6 stars
with accurate {\it Hipparcos} parallaxes, comparing the individual
parallaxes to the weighted mean value (5.77\,mas) using a $\chi^2$
test gives $\chi^2$/$\nu$ = 1.85/5 for a high $\chi^2$ probability of
87\%. For the full sample of {\it Hipparcos} and Tycho parallax
measurements (15 values), comparing the individual parallaxes to the
weighted mean value using a $\chi^2$ test gives $\chi^2$/$\nu$ =
11.7/14, and a $\chi^2$ probability of 63\%. The former is the more
demanding test, and suggests that the individual parallax measurements
(when available) are very consistent with the mean value. The later
test is less enlightening, as the the Tycho parallax errors are
typically large ($\sim$10 mas), and most B- and A-type field stars
will have actual parallaxes of $\sim$1-15 mas anyway, due to the
magnitude limit of the Tycho catalog. An upper limit on the dispersion
of distances will be presented in \S\ref{spread}.  Throughout the rest
of this paper, I adopt a group distance of 173$^{+16}_{-13}$ pc.

\subsection{Age \label{age}}

\begin{figure}
\epsscale{0.9}
\plotone{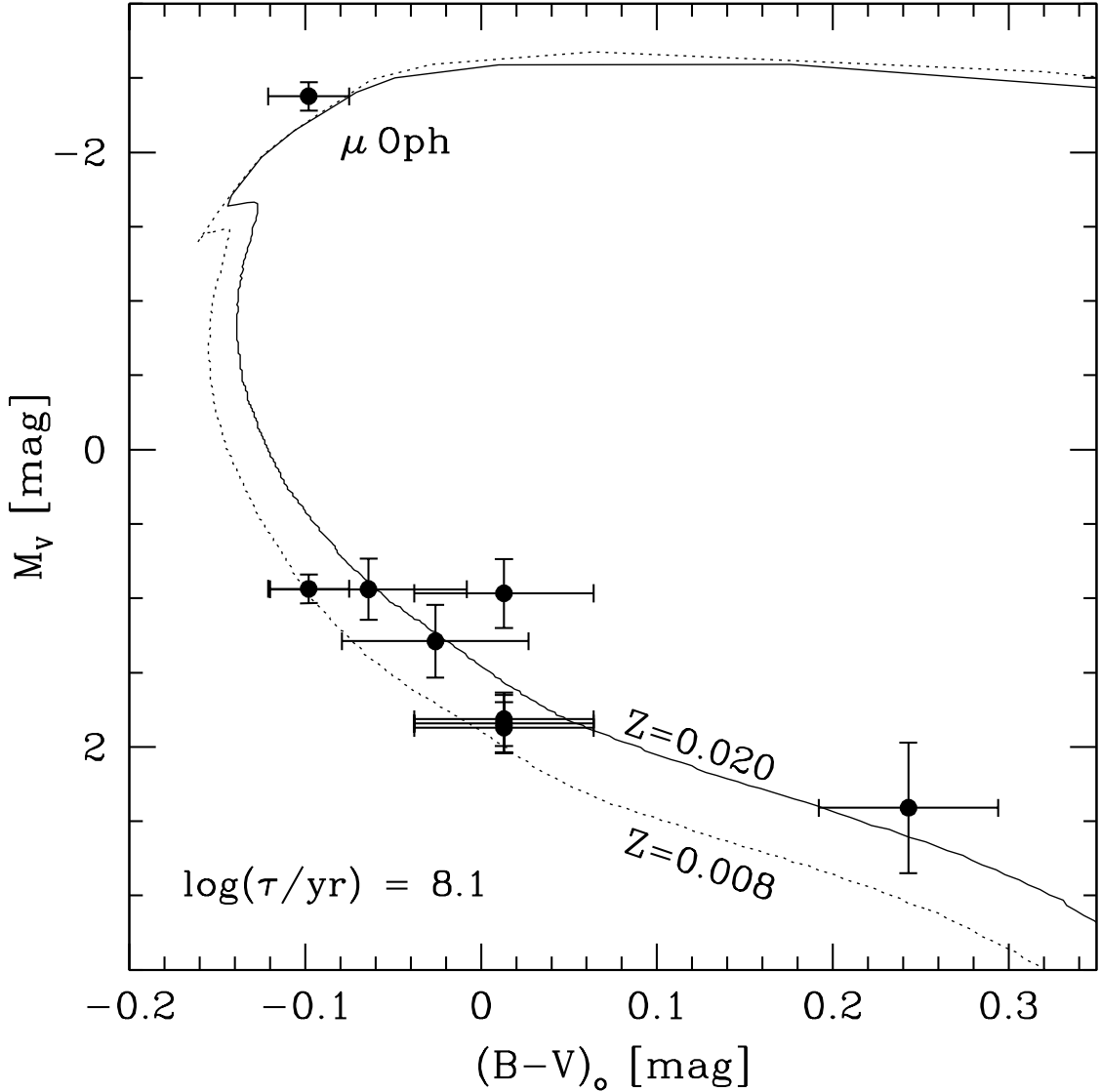}
\caption{ Dereddened color-magnitude diagram for $\mu$ Oph candidate
members selected via proper motions from Fig. \ref{fig:muoph_vector},
and using photometry from Table \ref{tab:muoph_members}. The errors in
\bvo\, reflect spectral type uncertainties of $\pm$1 subtype, whereas
the errors in \mv\, include errors in the color excess and \rv\ (but
not distance, which could introduce a systematic shift of $\pm$0.18
mag in \mv). Two isochrones are shown for log(age/yr) = 8.1, for two
metal mass fractions which bracket the solar metallicity ($Z$ $\simeq$
0.015). The isochronal age of $\mu$ Oph is nearly independent of
metallicity, and leads to a post-MS age estimate of 120\,$\pm$\,25 Myr
(\S\ref{age}).
\label{fig:muoph_bv}}
\end{figure}

With the reddening and distance towards the sample quantified, one can
attempt to estimate an isochronal age for the system. For this
purpose, I employ the evolutionary tracks and isochrones from
\citet{Lejeune01} for metal mass fractions of $Z$ = 0.008
and 0.020. The dereddened color vs. absolute magnitude diagram for the
candidate cluster sample is shown in Fig. \ref{fig:muoph_bv}.  The
metal fractions of the two sets of isochrones bracket the range of
recent estimates of the solar $Z$ \citep[0.0126 $<$ $Z_{\odot}$ $<$
0.0189; e.g.][]{Anders89,Grevesse98,Lodders03,Asplund04,Antia06}.  The
assumption of solar $Z$ should be an excellent first assumption, as
the local open cluster population has a mean metallicity consistent
with solar, with only a $\pm$0.1 dex dispersion in [Fe/H]
\citep{Twarog97}.  For the solar metal fraction ($Z$ $\simeq$ 0.015),
the isochronal age of the post-MS star $\mu$ Oph is log(age/yr) =
8.08\,$\pm$\,0.05, and nearly independent of $Z$ over the range
probed.  The main sequence stars are consistent (within their errors)
of lying on this isochrone for solar metallicity ($Z$ $\simeq$ 0.015)
at the mean {\it Hipparcos}-Tycho distance (173\,pc).  The turn-off
age estimate unfortunately hinges sensitively on the placement of the
post-MS (H shell-burning) star $\mu$ Oph, but Fig. \ref{fig:muoph_bv}
suggests that the color-magnitude-distance data for this kinematic
group are consistent with a coeval and co-distant cluster.

To test whether the age estimate for the $\mu$ Oph group is consistent
with that for other well-studied clusters, I estimated ages for the
$\alpha$ Per and Pleiades clusters using {\it Hipparcos} photometry of
evolved MS and post-MS stars. I adopt the distance and reddening for
the $\alpha$ Per cluster, as well as the reddening for the Pleiades,
from \citet{Pinsonneault98}, and use the recent Pleiades distance
estimate from \citet{Soderblom05}. Using the median dereddened CMD
position for the cluster members with $V$ $<$ 6$^m$ from
\citet{deZeeuw99}, I estimate a turnoff age of 95\,Myr for the
$\alpha$ Per cluster.  Using the same technique on a sample of high
mass stars in the Pleiades \citep[$V$ $<$ 5$^m$ members
from][]{Robichon99} yields an age of 130 Myr.  These turn off age
estimates are in excellent agreement ($\sim$5\%) with the most recent
Li-depletion boundary (LDB) ages of 85\,$\pm$\,10\,Myr ($\alpha$ Per)
and 130\,$\pm$\,20 Myr \citep[Pleiades;][]{Barrado04}.  Given the
statistical uncertainty in the isochronal age for $\mu$ Oph (12\%),
the cross-calibration error between the LDB and turnoff ages for the
benchmark $\alpha$ Per and Pleiades clusters ($\sim$5\%), and the
published uncertainties in the LDB ages \citep[here assumed to be
errors in the absolute ages; $\approx$10-15\%; ][]{Barrado04}, I
conservatively estimate the uncertainty in the age of the new cluster
to be $\approx$20\%. The age of the group is then 120\,$\pm$\,25 Myr.

\subsection{Space Motion \label{space_motion}}

The bulk space motion of the cluster is an interesting quantity to
compare to other nearby young stellar groups, and requires accurate
estimates of the group's distance, radial velocity, and proper
motion. The distance to the group (173$^{+16}_{-13}$ pc) was 
determined earlier (\S\ref{distance}), and appears to be well-constrained.
Unfortunately, at present, only two members of the new cluster have
radial velocity measurements, but they are consistent: $\mu$ Oph
\citep[--18.5\, $\pm$\, 1.4 \kms;][]{Barbier00} and HD 158450
\citep[--22.0\, $\pm$\, 4.2 \kms;][]{Grenier99}.  The weighted mean
radial velocity for these two stars is --18.9\,$\pm$\,1.3 \kms.  Using
the best available proper motions for the nine stars (Tycho-2 for
$\mu$ Oph, and UCAC2 for the remaining eight), I estimate a mean
proper motion of $\overline{\mu_{\alpha *}}$ = --12.1\, $\pm$\,
0.4 \masyr, and $\overline{\mu_{\delta}}$ = --20.6\, $\pm$\, 0.4 \masyr.
From these mean proper motion, radial velocity, and distance values, I
estimate the barycentric space motion of the group to be ($U, V, W$ =
--12, --24, --4 \kms), with $\pm$1\,\kms\, errors in each
component.  The space motion differs significantly from that of
components of the local star-forming complex \citep[the ``Gould
Belt''; $\lesssim$60 Myr; $U, V, W$ $\simeq$ --10, --14, --7 \kms;
e.g. Sco-Cen, Ori OB1, etc.;][]{Torra00}.
Despite its spatial proximity to Sco-Cen, the age and velocity of the
new group are completely inconsistent with it being a subgroup of the
nearest OB association. The new group is, however, surprisingly close
in both age {\it and} velocity to the 85 Myr-old $\alpha$ Per cluster
\citep[$U, V, W$ = --15, --26, --8 \kms;][]{Robichon99}, 130 Myr-old
Pleiades cluster ($U, V, W$ = --7, --28, --15 \kms) and 75-150 Myr-old
AB Dor moving group \citep[$U, V, W$ = --8, --26, --14
\kms;][]{Luhman05}. The $\mu$ Oph group and the three later groups all
lie within a few \kms\, of the ``Pleiades branch'' velocity-space
feature described by \ \citet{Skuljan99}. These clusters also share an
over-density in velocity-age space with early-type field stars
(``Group B4'') in the solar neighborhood with ages of
$\sim$150\,$\pm$\,50 Myr and mean velocity of \citep[$U, V, W$ =
--9$\pm$5, --26$\pm$3, --9$\pm$5 \kms;][]{Asiain99}. Ostensibly, the
``B4'' group and the AB Dor Moving Group may be the same entity (save
the slight physical clustering of young low-mass stars associated with
AB Dor itself).  Given the similarities in both age and space motion
between the $\mu$ Oph, Pleiades, and $\alpha$ Per clusters, and the AB
Dor and/or ``B4'' moving groups, it is conceivable that the newly
identified cluster formed in the same star-forming complex that
spawned these other well-known groups.

The properties of the new stellar aggregate are summarized in Table
\ref{tab:muoph_clus}.

\section{Discussion \label{discussion}}

\subsection{Solar Reflex Motion? \label{reflex_motion}}

Could this aggregate be composed of unrelated B- and A-type field
stars whose projected positions are clustered in the sky, but simply
demonstrating solar reflex motion? While solar reflex motion is
imparted on the observed proper motion of every star, the effect can
not be responsible for the nearly identical proper motions of the
eight early-type stars in the proposed aggregate.  The spread in
proper motions is an order of magnitude smaller than that expected for
an ensemble of young {\it field} stars (even if situated at identical
distances).

Suppose that the group is actually composed of field stars at similar
distances (170 pc) but with a distribution of 3D space motions
representative of B- and early A-type field stars. Here I simulate the
proper motions of early-type field stars using the velocity dispersion
tensor for stars with (\bv) $\in$ [-0.238, 0.139] from Table 1 of
\citet{Dehnen98}, using a simple Monte Carlo technique previously
employed by the author in \citet{Siegler03}. For the simulated data
sets, I adopt the solar peculiar motion with respect to the LSR from
\citet{Dehnen98}. I create a Monte Carlo sample of 10$^4$ stars with
$UVW$ vectors drawn from the \citet{Dehnen98} velocity dispersion
tensor, then assume that these stars lie at the distance and celestial
position of the center of the $\mu$ Oph aggregate, and calculate what
distribution of proper motions one would see. The simulated proper
motions are consistent with \mura\,\, =\, +3\,$\pm$\,8 (1$\sigma$)
\masyr\, and \mudec\, =\, --5\,$\pm$\,13 (1$\sigma$) \masyr.  The
predicted scatter in proper motion components ($\sigma_{\mu}$\,
$\simeq$\, 8--13\, \masyr) is roughly an order of magnitude larger
than the observed dispersion in the proper motion components for the
proposed cluster ($\sim$1\,\masyr), even before accounting for the
individual proper motion errors (mean errors $\sigma_{\mu}$ $\simeq$
1.0 \masyr; Table \ref{tab:muoph_astrom}).  The amount of {\it
intrinsic} velocity dispersion (unaccounted for by the observational
errors) is unresolvable (\S\ref{dynamics}) -- the proper motion errors
appear to account for all of the observed spread. The mean proper
motion of the proposed cluster is also not near the predicted proper
motion locus for the field population (which is within
$\sim$10\,\masyr\, of zero motion). From these calculations, one can
rule out solar reflex motion alone as the agent for the clumping of
the proper motions observed in Figs. \ref{fig:muoph_vector}.

\subsection{Dynamics \label{dynamics}}

How has the rather poor stellar nucleus of the $\mu$ Oph cluster
survived to an age of $\sim$10$^8$ years? The answer likely lies with
the high density of the nucleus. For the fiducial isochrone ($Z$ =
0.015; 120 Myr), the absolute magnitudes of the nucleus stars suggest
masses for the nucleus stars of 5.2\,\msun\, ($\mu$ Oph), 3.0\,\msun\,
(HD 159874), 2.8\,\msun\, (HD 160038), and 2.1\,\msun\, (HD
160037). The deprojected half-mass radius \citep{Spitzer87} for the
observed membership is approximately 0.4\,pc (corresponding closely to
the observed nucleus radius) and within this radius the stellar
density is at least $\simeq$49\,\msun\,pc$^{-3}$ \citep[600$\times$
the local Galactic disk density of $\sim$0.08
\msun\,pc$^{-3}$;][]{Creze98}.  This density is similar to that within
a similar size volume of the $\eta$ Cha cluster
\citep[$\simeq$56\,\msun\,pc$^{-3}$; calculated from Table 2 and
Fig. 7 of][]{Lyo04}. Using the formula of \citet{King62} and the Oort
constants from \citet{Feast97}, I estimate the cluster tidal radius
($r_t$) for the observed membership to be $\approx$4 pc.

Ideally one would like an estimate of the velocity dispersion to give
some indication of the dynamical state of the group (i.e. is it
bound?).  If the group were bound and conformed to a Plummer model,
then given the total mass (24\,\msun) and half-mass radius (0.4\,pc),
the central 1D velocity dispersion should be $\sim$0.25 \kms\,
\citep{Gunn88}. This is negligible compared to the proper motion
uncertainties, but consistent with the lack of a detectable spread in
the tangential motions. To test to see whether an intrinsic velocity
dispersion was present, I use the best long-baseline proper motions
available for each star (Tycho-2 for $\mu$ Oph, and 2UCAC for the
other seven stars; Table
\ref{tab:muoph_astrom}). The velocity dispersion can be estimated by
the distribution of proper motions in the $\tau$ (tangential)
direction, i.e. in the direction perpendicular to the great circle
joining the position of each star and the convergent point for the
group \citep[e.g.][and references therein]{Mamajek05}. The group
convergent point can be estimated from the space motion vector
($\alpha_{cvp}, \delta_{cvp}$ = 108$^{\circ}$.9, --30$^{\circ}$.7;
uncertainty of $\pm$2$^{\circ}$.7)\footnote{The classic convergent
point technique does not work well for stellar groups that are close
together, like the $\mu$ Oph group. The convergent point algorithm
described in \citet{Mamajek05} yields a group convergent point of
$\alpha_{cvp}, \delta_{cvp}$ $\simeq$
117$^{\circ}$($^{+81^{\circ}}_{-19^{\circ}}$),
--39$^{\circ}$($^{+24^{\circ}}_{-15^{\circ}}$); (68\% confidence
intervals), within a long narrow ($\approx$1$^{\circ}$.5 wide) error
ellipse oriented NW-SE. This convergent point is within 10$^{\circ}$
of the convergent point estimated from the space motion, which is
pleasing (and perhaps surprising) considering the size of the error
ellipse.}.  The mean proper motion of the eight stars towards the
convergent point is $\overline{\mu_{\upsilon}}$ = 23.9
\masyr\, (with 1$\sigma$ dispersion of 1.2 \masyr), and the mean
proper motion in the perpendicular direction is $\overline{\mu_{\tau}}$ = 0.0
\masyr\, (with 1$\sigma$ dispersion of 0.9 \masyr). The mean
uncertainties in the $\alpha$ and $\delta$ proper motion components
are both 1.0 \masyr, so the uncertainties in the rotated components
($\sigma_{\mu_{\upsilon}}$ and $\sigma_{\mu_{\tau}}$) are similarly
1.0 \masyr.  Given the observational proper motion uncertainties 
(1.0 \masyr), the dispersion in $\mu_{\tau}$ (0.9 \masyr) for these
nine stars is statistically consistent with parallel motion with
negligible (indeed {\it zero}) velocity dispersion. For reference, a
tangential velocity of 1\,\kms\,
\citep[typical for the velocity dispersion observed in OB association
subgroups; ][]{Briceno06} at $d$ = 173\,pc is equivalent to a proper
motion of 1.2 \masyr. A 1D velocity dispersion of 1\,\kms\, added in
quadrature to the proper motion uncertainties would have resulted in
an observed proper motion dispersion of 1.6\,\masyr, however this is
easily ruled out. Hence the data are consistent with the idea that the
velocity dispersion is negligible, but a value can not be derived from
modern astrometric data.

The evaporation time of the nucleus can be calculated following
\citet{Binney87} for two extreme cases. In the first case, where we
are dealing with a ``cluster remnant'' with only the four high mass
stars in the nucleus, the evaporation time is short ($\sim$10$^8$ yr).
This is perhaps not so surprisingly similar to the {\it isochronal}
age, given the ``poor'' appearance of the nucleus. In the other
extreme case, one posits that the four high-mass stars in the nucleus
represent a complete census of 2-5.5\,\msun\, members, and hypothesize
that there is an as-yet undetected entourage of low-mass members with
masses drawn from a \citet{Kroupa01} initial mass function. In this
case, the existence of 4 high-mass members implies that
$\sim$130\,$\pm$\,60 low-stars might exist\footnote{If the nine B-
and A-type stars represent a complete census of the 1.5-5.5\,\msun\,
population that initially formed, then assuming a \citet{Kroupa01}
IMF, the initial population may have been $\sim$200 systems. This is
similar in population to nearby ECs like NGC 2024, Mon R2, and Cha I
\citep{Porras03}.}, for a total nucleus mass of $\sim$60\,\msun.  In
this case the evaporation time is of order $\sim$400\,Myr. Reality
probably lies in between these two extreme cases. There are likely to
be some undetected low-mass members in the nucleus, but dynamical
evolution will have preferentially evaporated the low mass members
\citep[e.g.][]{delaFuenteMarcos95}. 

\subsection{ISM \label{ISM}}

In addition to the group's ``poorness'', the moderate extinction
(\av\, $\simeq$\, 0.9$^{m}$) towards the $\mu$ Oph group may be one
reason why it was previously unnoticed. There are 2 Lynds dark clouds
within 3$^{\circ}$ of $\mu$ Oph \citep[\#393, \#382;][]{Lynds62}. In
the new Catalog of Dark Clouds published by \citet{Dobashi05}, there
are 14 clouds and 22 ``clumps'' within 3$^{\circ}$ of $\mu$ Oph. The
low-opacity cloud LDN 393 which covers $\sim$20 square degrees
\citep{Lynds62} was broken down by \citet{Dobashi05} into $\sim$25
clumps, with clouds DUKSKUS H189, H159, and H177 accounting for most
of the cloud coverage (clouds are from their Table 7). The clumps
typically have peak extinctions of \av\, $\simeq$ 1-4\,mag. They are
likely in the foreground of the new cluster, and represent a part of
the Aquila Rift cloud complex connecting the Sco-Oph molecular clouds
\citep[$\ell$ $\simeq$ 355$^{\circ}$; $\sim$140\,pc;][]{deZeeuw99}
with the Serpens clouds \citep[$\ell$ $\simeq$ 30$^{\circ}$;
$\sim$225\,pc;][]{Straizys03}.  To estimate how much gas is associated
with the dark clouds near $\mu$ Oph, I take the integrated
extinction-area values from \citet{Dobashi05} for clouds within
3$^{\circ}$ (9\,pc, projected) of the cluster nucleus, assume a
distance of $d$ $\simeq$ 150\,pc, and adopt the dust-gas ratio from
\citet{Savage79}.  The total gas mass of the dark clouds is then
approximately 5000-13000 \msun.

\subsection{Spread in Distances \label{spread}}

As stated in \S\ref{distance}, the {\it Hipparcos} and Tycho
parallaxes for the nine members are statistically consistent with a
mean distance of $\simeq$173\,pc. However it is not clear whether they
are consistent with lying within a range of distances similar to their
projected size on the sky (i.e. $\sim$5-10\,pc), or the estimated
tidal radius (4\,pc). Here I estimate a plausible upper limit to the
spread in the distances to the individual cluster members, {\it
assuming that the members have identical space motions with negligible
velocity dispersion} (physically motivated by the results from
\S\ref{dynamics}).

A cluster member's distance from the Sun ($d_i$ in pc), angular
separation from its group convergent point ($\lambda_i$), proper
motion component pointing towards the group's convergent point
($\mu_{\upsilon i}$ in \masyr), and the magnitude of the cluster's
space velocity (i.e. $V$ in \kms), are related through the classic
formula \citep[e.g.][]{Atanasijevic71}:

\begin{equation}
d_i\,=\,\frac{1000\,V\,sin\,\lambda_i}{A\,\mu_{\upsilon i}}
\end{equation}

\noindent where A is the astronomical unit in useful units (4.74047
km\,yr\,s$^{-1}$). For cluster members with negligible velocity
dispersion and spread in $\lambda$, the predicted dispersion in the
distances amongst the cluster members ($\sigma^{int}_{d}$) is
approximately:

\begin{equation}
\sigma^{int}_d\,\simeq\,\frac{1000\,V\,sin\,\overline{\lambda}}{A\,\overline{\mu_{\upsilon}^2}}\,\sigma^{int}_{\mu_{\upsilon}}
\end{equation}

\noindent where I have defined $\sigma^{int}_{\mu_{\upsilon}}$ as the
intrinsic dispersion in $\mu_{\upsilon}$ values unaccounted for by
observational errors (here presumed to be due to the spread in
distances only), and $\overline{\lambda}$ = 133$^{\circ}$.9,
$\overline{\mu_{\upsilon}}$ = 23.9\,\masyr, and $V$ = 27.2\,\kms\, for
the cluster sample.  The observed 1$\sigma$ dispersion in
$\mu_{\upsilon}$ values is $\pm$1.2\,\masyr\ (\S\ref{dynamics}), and
the mean observational error in $\mu_{\upsilon}$ is $\pm$1.0\,\masyr.
This leaves an intrinsic scatter of $\sigma_{\mu_{\upsilon}}^{int}$
$\simeq$ 0.7\,\masyr\ of unaccounted dispersion. If we attribute this
spread in the $\mu_{\upsilon}$ proper motions solely to the dispersion
in distances to the cluster members using equations (4) and (5), then
the upper limit on the dispersion in distances is $\sigma^{int}_{d}$\,
$\simeq$\, 5\,pc. If all of the stars were at the same distance, and
the intrinsic spread in $\mu_{\upsilon}$ values were attributed to a
1D velocity dispersion, then the data would imply a velocity
dispersion of 0.6\,\kms. However, a 1D velocity dispersion this large
can be ruled out since the spread in $\mu_{\tau}$ values (the
component perpendicular to $\mu_{\upsilon}$) is negligible compared to
the observational errors (\S\ref{dynamics}). Based on this analysis,
it appears likely that the 1$\sigma$ dispersion in the distances to
the individual cluster members is of order $\pm$5\,pc, consistent with
the idea that the depth of the group is similar in size to the
projected diameter of the group, and similar in magnitude to the
estimated tidal radius.

\section{Summary \label{summary}}

A new, young, nearby candidate stellar aggregate in Ophiuchus is
reported. The group consists of the 4th magnitude B8 giant $\mu$ Oph
and eight co-moving B- and A-type stars in Ophiuchus. The
color-magnitude and astrometric data are consistent with the
hypothesis that these stars constitute a co-distant (173\,pc),
co-moving, and coeval (120\,$\pm$\,25\,Myr) stellar group.  The
group's one-dimensional velocity dispersion is unresolvable with the
best available long baseline proper motion data, consistent with the
minuscule value predicted from dynamical considerations
($\sim$0.2\,\kms). Roughly half of the group's observed stellar mass
($\approx$24\,\msun) appears to be concentrated within a nucleus with
half-mass radius of $r_h$ $\simeq$ 0.4\,pc. A kinematic analysis of
the proper motion components pointing towards the group convergent
point suggests that the intrinsic scatter in the distances to the
individual group members is probably of order $\pm$\,5\,pc, similar in
size to the inferred tidal radius (4\,pc). If the census of
1.5-5.5\,\msun\, members is complete, and the group formed with a
normal initial mass function, then the initial population of the new
group may have been of order $\sim$200 systems.  The space motion and
age of the group are similar to that of the Pleiades, $\alpha$ Per
cluster, and AB Dor Moving Group, suggesting that these entities may
have formed in the same complex some $\sim$10$^8$ yr ago.

Future observations are clearly desirable to first further test
the reality of the group, and secondly to enlarge and better
characterize its membership. Analysis of a photometric survey in the nuclear
region of the $\mu$ Oph cluster is underway (Kenworthy \& Mamajek, in
prep.) in hopes of identifying the low-mass members, and to further
constrain the cluster age via its pre-main sequence.  Low-resolution
optical spectra will then be used to confirm the youth of the low-mass
candidate members. High-resolution spectroscopy will be necessary to
measure the radial velocities of all of the candidates in order to
further constrain the membership, and independently estimate the
velocity dispersion of the group.

\acknowledgments

I would like to thank the referee, Massimo Robberto, for a very
thoughtful and timely review. The author is supported through a Clay
Postdoctoral Fellowship from the Smithsonian Astrophysical
Observatory. I would like to thank Matt Kenworthy and Nick Tothill for
their comments on an early draft.  This study used the Digitized Sky
Survey, SIMBAD, and Vizier services.

\clearpage 

\begin{deluxetable}{llrrcl}
\tablewidth{0pt}
\tabletypesize{\scriptsize}
\tablecaption{Astrometric Data for Candidate Members of $\mu$ Oph Cluster \label{tab:muoph_astrom}}
\tablehead{\colhead{Star} & \colhead{Astrometric} & \colhead{\mura} & \colhead{\mudec} & \colhead{$\varpi$} & \colhead{2MASS}\\
\colhead{Name} & \colhead{Alias} & \colhead{(\masyr)} & \colhead{(\masyr)} & \colhead{(mas)} & \colhead{J}}
\startdata
HD 158450&HIP 85618      & -14.1\,$\pm$\,2.6&-21.2\,$\pm$\,1.2 &  6.04\,$\pm$\,2.15 &17294397-0801031\\
HD 158450&TYC 5659-196-1 & -13.4\,$\pm$\,1.1&-23.9\,$\pm$\,1.2 & -5.40\,$\pm$\,9.80 &''\\
HD 158450&2UCAC 28939300 & -13.3\,$\pm$\,1.1&-21.6\,$\pm$\,1.1 &  ...               &''\\
HD 158450&ASCC 1321674   & -12.8\,$\pm$\,1.7&-23.7\,$\pm$\,1.6 &  5.51\,$\pm$\,2.15 &''\\
HD 158838&TYC 5663-310-1 & -10.5\,$\pm$\,1.4&-22.1\,$\pm$\,1.5 &  0.50\,$\pm$\,6.90 &17315378-1027164\\
HD 158838&2UCAC 28223894 & -10.3\,$\pm$\,1.2&-21.9\,$\pm$\,1.1 &  ...               &''\\
HD 158838&ASCC 1404541   &  -9.9\,$\pm$\,1.9&-21.5\,$\pm$\,1.3 &  0.50\,$\pm$\,6.90 &''\\
HD 158875&TYC 5659-186-1 & -12.0\,$\pm$\,1.3&-19.5\,$\pm$\,1.5 & 13.40\,$\pm$\,25.60&17320307-0912018\\
HD 158875&ASCC 1321733   & -11.2\,$\pm$\,1.2&-19.5\,$\pm$\,2.1 & 13.40\,$\pm$\,25.60&''\\ 
HD 158875&2UCAC 28575500 & -12.7\,$\pm$\,1.0&-21.1\,$\pm$\,1.2 &  ...               &''\\
HD 159209&HIP 85936      &  -9.1\,$\pm$\,1.4&-20.7\,$\pm$\,0.5 &  4.58\,$\pm$\,1.40 &17334689-0755034\\
HD 159209&TYC 5659-41-1  & -11.0\,$\pm$\,1.3&-20.8\,$\pm$\,1.3 &  2.60\,$\pm$\,11.80&''\\
HD 159209&2UCAC 29123893 & -13.1\,$\pm$\,1.1&-21.2\,$\pm$\,1.0 &  ...               &''\\
HD 159209&ASCC 1321774   &  -9.5\,$\pm$\,1.6&-21.2\,$\pm$\,0.6 &  4.55\,$\pm$\,1.39 &''\\
HD 159874&HIP 86240      & -11.5\,$\pm$\,1.0&-21.4\,$\pm$\,0.6 &  6.74\,$\pm$\,1.05 &17372392-0802125\\
HD 159874&TYC 5660-204-1 & -12.0\,$\pm$\,1.2&-19.2\,$\pm$\,1.2 & -3.20\,$\pm$\,7.30 &''\\
HD 159874&2UCAC 28940598 & -12.5\,$\pm$\,1.0&-19.8\,$\pm$\,0.9 &  ...               &''\\
HD 159874&ASCC 1321859   & -12.0\,$\pm$\,1.3&-20.5\,$\pm$\,0.9 &  6.53\,$\pm$\,1.04 &''\\
$\mu$ Oph&HIP 86284      & -11.7\,$\pm$\,0.7&-20.4\,$\pm$\,0.5 &  5.94\,$\pm$\,0.81 &17375071-0807075\\
$\mu$ Oph&TYC 5660-589-1 & -11.4\,$\pm$\,0.9&-19.1\,$\pm$\,0.8 &  2.60\,$\pm$\,3.00 &''\\
$\mu$ Oph&ASCC 1321867   & -11.3\,$\pm$\,1.0&-20.2\,$\pm$\,0.8 &  5.71\,$\pm$\,0.81 &''\\
HD 160038&TYC 5660-52-1  & -14.7\,$\pm$\,1.3&-20.9\,$\pm$\,1.3 &  3.70\,$\pm$\,6.60 &17380907-0808034\\
HD 160038&2UCAC 28940731 & -13.3\,$\pm$\,1.0&-21.0\,$\pm$\,1.1 &  ...               &''\\
HD 160038&ASCC 1321875   & -13.7\,$\pm$\,1.6&-20.6\,$\pm$\,1.7 &  3.70\,$\pm$\,6.59 &''\\
HD 160037&HIP 86318      & -10.7\,$\pm$\,1.2&-20.8\,$\pm$\,0.8 &  6.21\,$\pm$\,1.23 &17381246-0806212\\
HD 160037&TYC 5660-255-1 & -10.5\,$\pm$\,1.1&-20.2\,$\pm$\,1.1 &-18.90\,$\pm$\,11.80&''\\
HD 160037&2UCAC 28940741 & -11.3\,$\pm$\,0.9&-20.7\,$\pm$\,0.9 &  ...               &''\\
HD 160037&ASCC 1321877   &  -9.6\,$\pm$\,1.0&-20.5\,$\pm$\,1.0 &  5.94\,$\pm$\,1.23 &''\\
HD 160142&HIP 86343      & -11.7\,$\pm$\,1.5&-20.3\,$\pm$\,1.0 &  6.26\,$\pm$\,1.66 &17383811-0850277\\
HD 160142&TYC 5660-504-1 & -11.6\,$\pm$\,1.2&-19.0\,$\pm$\,1.2 & -4.80\,$\pm$\,12.60&''\\
HD 160142&2UCAC 28756716 & -11.0\,$\pm$\,0.9&-18.8\,$\pm$\,1.0 &  ...               &''\\
HD 160142&ASCC 1321889   & -11.4\,$\pm$\,1.6&-20.3\,$\pm$\,1.4 &  6.07\,$\pm$\,1.65 &''\\
\enddata
\tablecomments{The astrometric aliases provide the reference for the
proper motion and parallax data, which are from the {\it Hipparcos}
and Tycho catalog \citep{Perryman97}, Tycho-2 catalog \citep{Hog00},
2UCAC catalog \citep{Zacharias04}, and the ASCC compiled catalog
\citep{Kharchenko01}. For stars with TYC names, the proper motions are
from Tycho-2 \citep{Hog00} and the parallaxes are from Tycho-1
\citep{Perryman97}.}
\end{deluxetable}

\clearpage

\begin{deluxetable}{lllcccccccccccl}
\tablewidth{0pt}
\rotate
\tabletypesize{\scriptsize}
\tablecaption{Stellar Data for Candidate Members of the $\mu$ Oph Cluster\label{tab:muoph_members}}
\tablehead{\colhead{\#}&\colhead{Name}&\colhead{Spec.}&\colhead{$V$}  &\colhead{\bv}  &\colhead{\bv} &\colhead{$J$}  &\colhead{$H$}  &\colhead{$K_s$}&\colhead{\ebv} &\colhead{\evj} &\colhead{\evh} &\colhead{\evk} &\colhead{\rv}&\colhead{Notes}\\
           \colhead{}  &\colhead{}    &\colhead{Type} &\colhead{(mag)}&\colhead{(mag)}&\colhead{Ref.}&\colhead{(mag)}&\colhead{(mag)}&\colhead{(mag)}&\colhead{(mag)}&\colhead{(mag)}&\colhead{(mag)}&\colhead{(mag)}&\colhead{}   &\colhead{}}
\startdata
1 & $\mu$ Oph & B8 II-IIImnp & 4.58 & 0.12 & 1,2,3,4,5 & 4.4: & 4.2: & 4.2: & 0.22 & \nodata & \nodata & \nodata &\nodata& (a)\\
2 & HD 159874 & B9 IV/V      & 7.83 & 0.11 & 1         & 7.45 & 7.42 & 7.39 & 0.21 & 0.47    & 0.59    & 0.63    &3.3    & \nodata \\
3 & HD 160038 & B9 V         & 7.98 & 0.18 & 1         & 7.44 & 7.40 & 7.36 & 0.24 & 0.60    & 0.70    & 0.73    &3.5    & (b)\\
4 & HD 160037 & A0 V         & 8.90 & 0.31 & 1         & 8.20 & 8.11 & 8.08 & 0.30 & 0.65    & 0.77    & 0.76    &3.0    & \nodata \\
5 & HD 158838 & B9.5 V       & 8.30 & 0.16 & 1         & 7.72 & 7.74 & 7.63 & 0.19 & 0.60    & 0.63    & 0.69    &4.3    & \nodata \\
6 & HD 158450 & Ap Si(Cr)    & 8.55 & 0.37 & 1,3       & 7.59 & 7.54 & 7.41 & 0.31 & 0.90    & 0.98    & 0.98    &3.9    & (c)\\
7 & HD 160142 & A0 V         & 8.99 & 0.31 & 1         & 8.26 & 8.21 & 8.13 & 0.29 & 0.68    & 0.76    & 0.80    &3.2    & \nodata \\
8 & HD 159209 & A0 V         & 9.00 & 0.29 & 1         & 8.25 & 8.20 & 8.10 & 0.28 & 0.70    & 0.78    & 0.83    &3.5    & \nodata \\
9 & HD 158875 & A(8) (p Si)  &10.16 & 0.51 & 1         & 8.65 & 8.34 & 8.21 & 0.28 & 1.16    & 1.34    & 1.39    &6.1    & (d)\\
\tableline
mean   &\nodata & \nodata   &\nodata&\nodata&\nodata&\nodata&\nodata&\nodata& 0.26 & 0.65    & 0.74    & 0.77    &3.5    & (e) \\
s.e.m. &\nodata & \nodata   &\nodata&\nodata&\nodata&\nodata&\nodata&\nodata& 0.02 & 0.05    & 0.05    & 0.04    &0.2    & (e) \\
st.dev.&\nodata & \nodata   &\nodata&\nodata&\nodata&\nodata&\nodata&\nodata& 0.05 & 0.13    & 0.13    & 0.11    &0.5    & (e) \\
\enddata
\tablecomments{All spectral types are from \citet{Houk99}, $V$
magnitudes are from \citet{Perryman97}, and $JHK$ photometry is from
2MASS \citep{Cutri03}. Reference numbers for the best estimate
of the \bv\, color are: (1) \citet{Perryman97},
(2) \citet{Johnson66}, (3) \citet{Hauck98}, where $b-y$, $m1$, and
$c1$ colors were converted to Johnson \bv\, via 
\citet{Turner90}, (4) \citet{Cousins64},
(5) \citet{Crawford63}. The uncertainties in the \rv\, values for each star are $\approx$0.2.
The last three rows are the unweighted mean, standard error of the mean, and standard deviation for the
sample.}
\tablenotetext{a}{The 2MASS photometry for $\mu$ Oph is very poor ($\sim$0.2-0.3 mag
errors per band), so I do not use this data to constrain the near-IR color excesses or \rv.}
\tablenotetext{b}{
{\it ROSAT} FSC X-ray source \citep[1RXS 173809.5-080749;][]{Voges00}
with flux rate of 1.80e-2 ct\,s$^{-1}$. At distance of 173\,pc this
translates to an X-ray luminosity of log\,($L_X$/erg\,s$^{-1}$)\, =\, 29.8
and fractional X-ray/bolometric luminosity ratio of
log\,$L_{X} / L_{bol}$\, =\, --5.4.  The X-ray flux is probably
from an active low-mass companion, as the X-ray luminosity is very
similar to that of low-mass stars seen in other $\sim$100\,Myr groups
\citep[e.g. Pleiades;][]{Stauffer94}.}  

\tablenotetext{c}{HD 158450 is a 0''.4 separation binary with a faint companion
\citep[$\Delta V$ = 2.0;][]{Worley96}. For the reddening calculations,
I assumed photospheric colors appropriate for an A0 star
\citep{Grenier99}.}

\tablenotetext{d}{HD 158875 is a 0''.9 separation binary with a faint companion
\citep[$\Delta V$ = 3.0;][]{Worley96}.}

\tablenotetext{e}{HD 158875 is omitted from inclusion in calculating the mean, standard error,
and standard deviation of \rv (see \S\ref{reddening}).}

\end{deluxetable}

\clearpage

\begin{deluxetable}{ll}
\tablewidth{0pt}
\tablecaption{Properties of the $\mu$ Oph Cluster (Mamajek 2)\label{tab:muoph_clus}}
\tablehead{\colhead{Property} & \colhead{Quantity}}
\startdata
Lucida                    & $\mu$ Oph ($V$\,=\,4.58$^m$, B8II-IIImnp)\\
Nucleus Center (J2000)    & 17$^h$38$^m$ --8$^{\circ}$ 06'\, (\S\ref{analysis})\\
Nucleus Center (Galactic) & 17$^{\circ}$.0, +12$^{\circ}$.3\, (\S\ref{analysis})\\
Angular Diam. of Nucleus  & 20' (\S\ref{analysis})\\
Physical Diam. of Nucleus & 1\,pc (\S\ref{analysis},\S\ref{distance})\\
Parallax ($\overline{\varpi}$)& 5.77\,$\pm$\,0.48 mas\, (\S\ref{distance})\\
Distance ($\overline{D}$)  & 173$^{+16}_{-13}$ pc\, (\S\ref{distance})\\
$\overline{(m-M)_o}$       & 6.19$^{+0.17}_{-0.19}$ mag\, (\S\ref{distance})\\
$\overline{\mu_{\alpha *}}$& --12.1\,$\pm$\,0.4 \masyr\, (\S\ref{space_motion})\\
$\overline{\mu_{\delta}}$  & --20.6\,$\pm$\,0.4 \masyr\, (\S\ref{space_motion})\\
$\overline{v_{rad}}$       & --18.9\,$\pm$\,1.3 \kms\, (\S\ref{space_motion})\\
Barycentric Vel. ($U,V,W$) & (--12, --24, --4)\,$\pm$\,(1, 1, 1) \kms\, (\S\ref{space_motion})\\
$\overline{E_{B-V}}$       & 0.26\,$\pm$\,0.02\,mag\, (\S\ref{reddening})\\
$\overline{R_V}$ (= \av/\ebv) & 3.5\,$\pm$\,0.2\, (\S\ref{reddening})\\
Mass of Candidate Members & 24\,\msun\, (\S\ref{dynamics})\\
Half-Mass Radius ($r_h$)  & 0.4\,pc\, (\S\ref{dynamics})\\
Tidal Radius ($r_{t}$)    & 4\,pc\, (\S\ref{dynamics})\\
Age                       & 120\,$\pm$\,25 Myr\, (\S\ref{age})\\
Absolute Magnitude (\mv)  & --2.5$^m$\\
\enddata
\end{deluxetable}

\end{document}